\pdfoutput=1
\documentclass[reprint, aps, prd, preprintnumbers, amsmath, amssymb, superscriptaddress, nofootinbib]{revtex4-1}

\usepackage{amssymb,amsmath,bbold,slashed}
\usepackage{graphicx,soul}
\usepackage[usenames,dvipsnames]{xcolor}
\usepackage[unicode=true,pdfusetitle,
 bookmarks=true,bookmarksnumbered=false,bookmarksopen=false,citecolor=Turquoise,
 breaklinks=false,pdfborder={0 0 1},backref=false,colorlinks=true,pdfpagemode=FullScreen]
 {hyperref}

\begin{document}

\preprint{IPMU18-0076}

\title{Muon $g-2$ and Dark Matter in the MSSM}

\author{Peter Cox}
\email[]{peter.cox@ipmu.jp}
\affiliation{Kavli IPMU (WPI), UTIAS, University of Tokyo, Kashiwa, Chiba 277-8583, Japan}
\author{Chengcheng Han}
\email[]{chengcheng.han@ipmu.jp }
\affiliation{Kavli IPMU (WPI), UTIAS, University of Tokyo, Kashiwa, Chiba 277-8583, Japan}
\author{Tsutomu T. Yanagida}
\email[]{tsutomu.tyanagida@ipmu.jp \\ Hamamatsu Professor}
\affiliation{Kavli IPMU (WPI), UTIAS, University of Tokyo, Kashiwa, Chiba 277-8583, Japan}

\begin{abstract}
We investigate the possibility that both dark matter and the long-standing discrepancy in the anomalous magnetic moment of the muon may be explained within the MSSM. 
In light of the stringent bounds from direct detection, we argue that the most promising viable scenarios have bino-like dark matter produced via either bino-wino or bino-slepton co-annihilation. 
We find that the combination of next-generation direct detection experiments and the LHC will be able to probe much of the interesting parameter space, however a future high-energy collider is needed to comprehensively explore this scenario. 
\end{abstract}

\maketitle

\section{Introduction}
Supersymmetry remains one of the most attractive frameworks for physics beyond the Standard Model. 
While its absence so far at the LHC is in some tension with our ideas about naturalness, the compelling motivations for low-scale supersymmetry essentially remain intact. 
Alongside naturalness and grand unification, one of these motivations is the natural presence of a dark matter (DM) candidate whose stability can be guaranteed by $R$-parity; in the case of the MSSM this is the lightest neutralino. 
A further interesting motivation lies in the long standing disagreement between the experimental measurement of the anomalous magnetic moment of the muon~\cite{hep-ex/0602035} and its SM prediction~\cite{1311.2198}; this discrepancy can be explained with relative ease within the MSSM. 

Given the recent progress and future sensitivity of DM searches and a new, higher precision measurement of the muon $g-2$ coming in the near future~\cite{1701.02807}, it is an interesting time to revisit the possibility that both DM and the muon $g-2$ could be simultaneously explained within the MSSM. 
Such a possibility has of course been widely considered in the literature~\cite{1404.4841, 1406.6925, 1409.3930, 1503.08219, 1504.00505, 1505.05877, 1505.05896, 1507.01395, 1608.03641, 1704.05287, 1710.11091}, although often using large scans over parameter space where the relevant physics in the surviving regions can be somewhat obscured. 
Furthermore, the requirement that the DM saturates the observed abundance, while satisfying the bounds from direct detection, is not always taken into account. 
In this work we aim to explore in detail the current status of this scenario in light of the latest experimental results, and discuss the prospects for testing it in the future. 
Given the existing constraints, we argue that the most promising regions of parameter space are those with a bino-like LSP, where the DM abundance is obtained via bino-wino or bino-slepton co-annihilation.
As we shall see, a complimentary approach involving DM direct detection and the HL-LHC will be able to probe significant regions of this parameter space, while a future multi-TeV lepton collider would be able to comprehensively explore this scenario.

\section{Dark matter and $(g-2)_\mu$ in the MSSM}

\subsection{Muon $g-2$}
The anomalous magnetic moment of the muon provides a sensitive probe of new physics, due to the high level of precision attained in both the experimental measurement~\cite{hep-ex/0602035} and the SM prediction~\cite{1311.2198}. 
The well-known discrepancy between theory and experiment, which is of the same order as the EW contributions, is given by~\cite{Patrignani:2016xqp}
\begin{equation}
  a_\mu^\text{exp} - a_\mu^\text{SM} = \left(2.68 \pm 0.63 \pm 0.43\right) \times 10^{-9} \,,
\end{equation}
where the errors correspond to the experimental measurement and theoretical prediction respectively, leading to a deviation of around $3.5\sigma$. 
This discrepancy should be further clarified in the near future by the E989 experiment at Fermilab, which aims to reduce the experimental uncertainty by a factor of four~\cite{1701.02807}. 

Within the MSSM, the one-loop contributions to the muon $g-2$ arise from diagrams with a chargino-sneutrino or neutralino-smuon loop~\cite{hep-ph/9512396, hep-ph/9308336, hep-ph/9507386}. 
An explanation of the current discrepancy therefore generally requires relatively light sparticles in the EW sector:
\begin{equation}
  |a_\mu^\text{SUSY}| \approx 1.3 \times 10^{-9} \left(\frac{100\,\text{GeV}}{M_{\text{SUSY}}}\right)^2 \tan\beta\,,
\end{equation}
although the relevant mass scales can be increased by taking large $\tan\beta$.

\begin{figure}[ht]
  \includegraphics[width=0.3\textwidth]{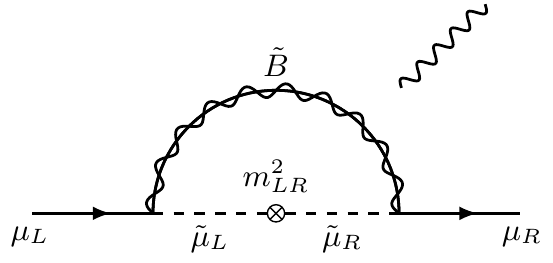}
  \caption{Bino-slepton contribution to the muon $g-2$.}
  \label{fig:bino-slepton}
\end{figure}

One contribution that will play a particularly important role in the following sections is the $\tilde{B}$-$\tilde{\mu}_L$-$\tilde{\mu}_R$ loop shown in Fig.~\ref{fig:bino-slepton}. 
This contribution is proportional to the left-right smuon mixing, $m_{LR}^2$, and so can receive a significant enhancement in the large $\mu\tan\beta$ limit (we will always assume the $A$-term is negligible). 
As we shall show, a relatively large $\mu$ is also favoured from the point of view of DM, in order to evade the strong bounds from direct detection. 

In our analysis we also include the leading two-loop contributions. 
The first of these is the log-enhanced QED correction, which can be considered as the renormalisation group evolution of the effective operator down to the muon mass scale, and generally leads to a reduction of around 10\%~\cite{hep-ph/9803384}. 
The other important two-loop effects are those proportional to $\tan^2\beta$~\cite{0808.1530}, which become increasingly important for large $\mu\tan\beta$. 
These arise from $\tan\beta$-enhanced diagrams that give corrections to the muon Yukawa coupling, and have been resummed to all orders~\cite{hep-ph/9912516}. 
Including these effects, the SUSY contribution to the $g-2$ is given by
\begin{equation} \label{eq:Delta}
  a_\mu^\text{SUSY} = \left(1-\frac{4\alpha}{\pi}\ln\frac{m_{\tilde{\mu}}}{m_\mu}\right) \left(\frac{1}{1+\Delta_\mu}\right) a_\mu^\text{1-loop} \,,
\end{equation}
where $\Delta_\mu\propto\mu\tan\beta$, with the full expression given in Ref.~\cite{0808.1530}. 
For a discussion of other known two-loop contributions see Ref.~\cite{1510.08071}. 

\subsection{Dark Matter}
The lightest neutralino in the MSSM provides a natural WIMP DM candidate. 
We shall assume that its abundance is determined entirely via thermal freeze-out and require that this saturates the observed DM density. 
Perhaps the simplest possibilities are (almost) pure higgsino or pure wino DM; however, in both cases the masses required to obtain the correct relic density ($\sim1\,$TeV~\cite{0706.4071} and $\sim3\,$TeV~\cite{hep-ph/0610249} respectively) are too large to allow for a simultaneous explanation of the muon $g-2$. 
The next logical choice is the 'well-tempered' neutralino where the LSP is made up of some $\tilde{B}-\tilde{H}-\tilde{W}$ admixture in order to achieve the correct relic abundance. 
This generically leads to a large DM-nucleon scattering cross-section and as such is now strongly constrained by DM direct detection experiments~\cite{1609.06735, 1701.02737, 1701.05869}.\footnote{These constraints are not applicable to the case of gauge mediation~\cite{1805.01607}, since the LSP is most likely the gravitino.} 

There are of course ways to avoid these constraints, such as the 'blind-spot' regions~\cite{1211.4873, 1612.02387} where the spin-independent and/or spin-dependent scattering cross-sections vanish at tree-level. 
We will not consider these special cases in detail, but it is worth noting that small values of $\tan\beta$ are required in order to satisfy both the blind-spot condition and the relic density, potentially making it more difficult to also explain the muon $g-2$. 
If the additional Higgs bosons are not too heavy, there can also be destructive interference between the $h/H$ contributions to the spin-independent scattering cross-section~\cite{1404.0392}; however, this possibility is beginning to come into tension with LHC searches for the pseudoscalar Higgs~\cite{1709.07242, 1803.06553}. 
Lastly, the $Z/h/A$-funnel regions~\cite{1303.3040, hep-ph/0106275} can also evade the current direct detection bounds, although the $Z$ and $h$ funnels are expected to be probed in the future~\cite{1603.07387}.

With the above caveats about certain special regions of parameter space, the strong bounds from direct detection therefore motivate us to focus on a bino-like LSP. 
Such an LSP is generally over-abundant, however co-annihilations can be used to enhance the annihilation and achieve the correct relic abundance. 
The annihilation rate today is then also expected to be well below the current sensitivity of indirect detection experiments. 
Note that the strong bounds on squarks and gluinos from LHC searches~\cite{1704.07781, 1705.04673, 1708.08232, 1712.02332} suggest that these should be somewhat heavier than the electroweak sector particles involved in explaining the muon $g-2$; we shall therefore assume that they are decoupled and play no important role in the DM phenomenology. 
This then leaves us with just two possibilities to explore: $\tilde{B}-\tilde{W}$ and $\tilde{B}-\tilde{l}$ co-annihilation\footnote{The importance of these two scenarios was also identified in a recent likelihood analysis of the pMSSM11~\cite{1710.11091}.}.

\section{Current Status and Future Prospects}

We now turn to the two scenarios of interest in detail. 
To begin with let us explicitly state our assumptions. 
We assume that all coloured sparticles are decoupled, motivated by the negative searches at the LHC~\cite{1704.07781, 1705.04673, 1708.08232, 1712.02332}. 
Similarly, we impose that the additional Higgs bosons are heavy, as suggested by LHC searches (at least for large $\tan\beta$)~\cite{1709.07242, 1803.06553}. 
In other words, we allow only for the presence of light neutralinos, charginos, and first and second generation sleptons (we also assume $m_{\tilde{l}}\equiv m_{\tilde{e}_L}=m_{\tilde{e}_R}=m_{\tilde{\mu}_L}=m_{\tilde{\mu}_R}$); we will also comment below on the implications of light staus. 
For concreteness, all other sparticles are assumed to have masses $\sim3$\,TeV, which is also the scale at which the soft masses are defined, and $A_t$ is fixed to obtain a Higgs mass of 125\,GeV. 
We use the spectrum generator {\tt SuSpect}~\cite{hep-ph/0211331} with {\tt MicrOMEGAs-4.3.5}~\cite{1606.03834} to calculate the relic abundance and DM-nucleon scattering cross-section, and the 1-loop SUSY contributions to the $g-2$.

\subsection{$\tilde{B}-\tilde{W}$ co-annihilation}

\begin{figure*}[t]
  \includegraphics[width=0.45\textwidth]{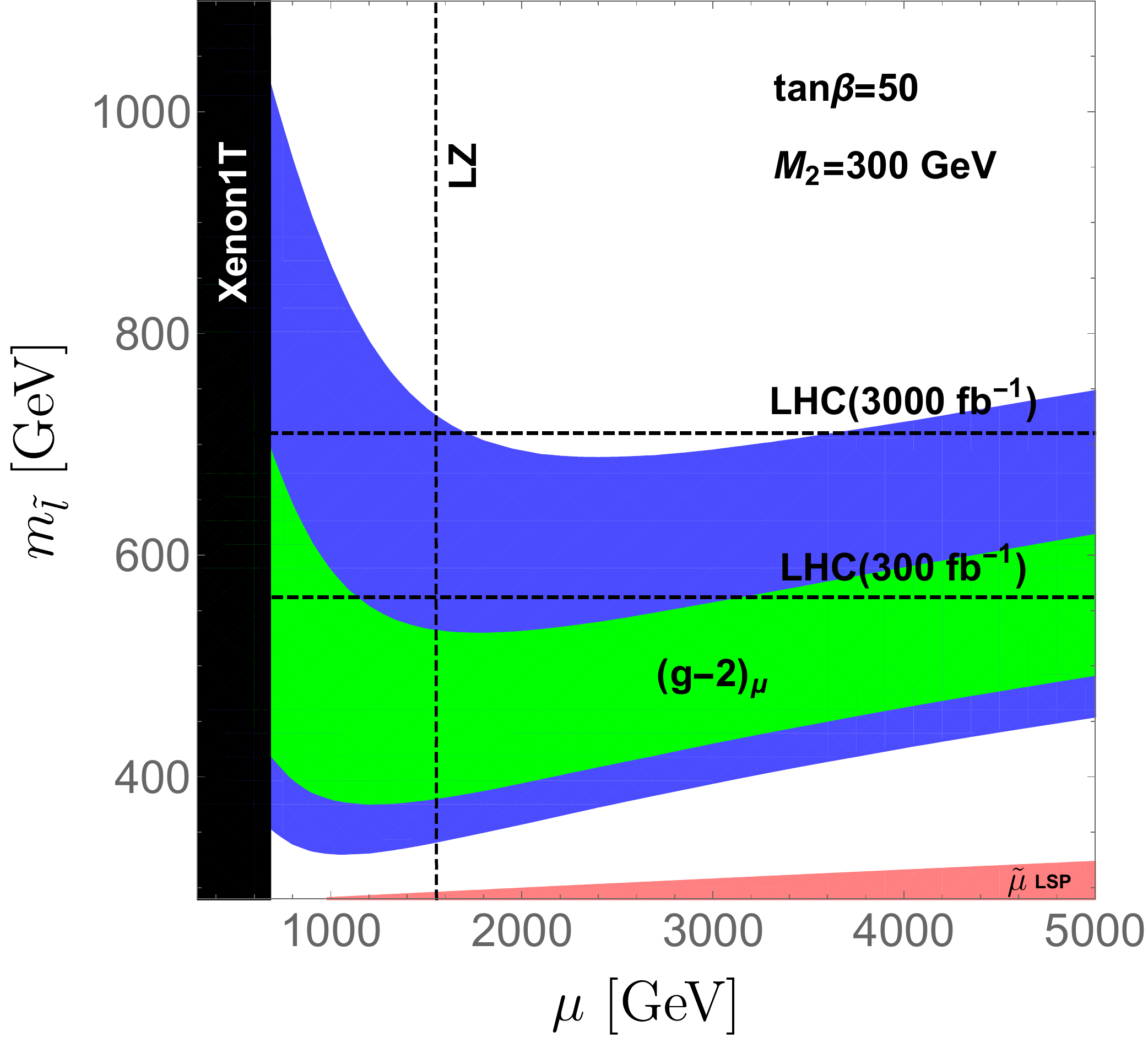}
  \includegraphics[width=0.45\textwidth]{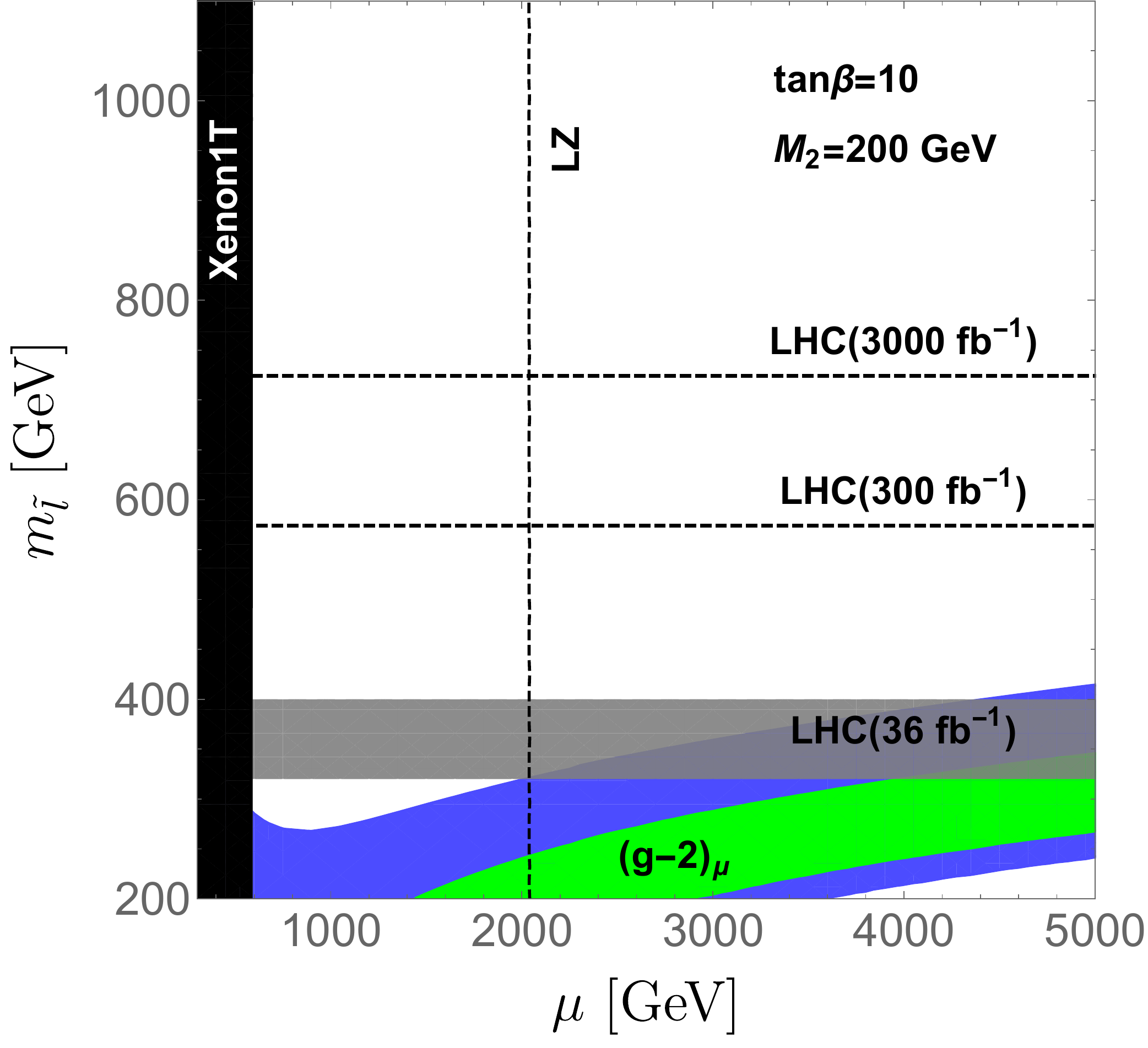}
  \caption{Current constraints and future projections for the case of $\tilde{B}-\tilde{W}$ co-annihilation. The green\,(blue) regions can explain the observed $(g-2)_\mu$ at $1\sigma\,(2\sigma)$. The black and grey regions are excluded by Xenon-1T~\cite{1705.06655} and the LHC~\cite{1803.02762} respectively, while the dashed lines denote projected future reach. In the pink region the lightest smuon becomes the LSP. The left\,(right) panels are for $\tan\beta=50\,(10)$ and $M_2=(300)\,200$\,GeV, and we have fixed $M_2-M_1=15$\,GeV.}
  \label{fig:B-W}
\end{figure*}

Obtaining the observed DM density through $\tilde{B}-\tilde{W}$ co-annihilation requires an $\mathcal{O}(10)\,$GeV mass-splitting between the bino-like LSP and wino-like NLSP~\cite{1311.2162, 1403.0715}. 
The precise mass-splitting required depends on both the LSP mass and, to a lesser extent, the bino-wino mixing. 
The compressed spectrum makes this scenario somewhat challenging to directly probe at colliders. 
Nevertheless, ATLAS~\cite{1712.08119} and CMS~\cite{CMS-PAS-SUS-16-025} have performed dedicated searches for such compressed spectra. 
The latest CMS search with $12.9\,\text{fb}^{-1}$ at $\sqrt{s}=13$\,TeV sets a limit of $m_{\tilde{\chi}^0_2/\tilde{\chi}^\pm_1}>195$\,GeV for $\Delta m=20$\,GeV~\cite{CMS-PAS-SUS-16-025}. 
This assumes that $\tilde{\chi}^0_2/\tilde{\chi}^\pm$ decay via off-shell gauge bosons; however, decays can also be mediated by the light sleptons, leading to a stronger limit in some regions of parameter space. 

In Fig.~\ref{fig:B-W}, we show the regions of parameter space which can explain the muon $g-2$, along with the latest experimental constraints. 
In the left (right) panel we have taken $\tan\beta=50\,(10)$, and $M_2=300\,(200)$\,GeV which is sufficient to satisfy the bounds from the compressed chargino searches in the regions of parameter space relevant for the $g-2$. 
We have also fixed $M_2-M_1=15$\,GeV in order to obtain the correct relic density\footnote{The precise value of $M_1$ required also has a mild $\mu$ dependence; however this has a negligible effect on our results.}. 
Let us focus initially on the left panel. 
The regions consistent with the $g-2$ measurement at $1\sigma\,(2\sigma)$ are shown in green\,(blue). 
For large $\mu$ the $\tilde{B}$-$\tilde{\mu}_L$-$\tilde{\mu}_R$ contribution in Fig.~\ref{fig:bino-slepton} dominates.
On the other hand, for $\mu\lesssim1.5$\,TeV the chargino contribution starts to become important, and the best-fit region moves towards larger slepton masses. 
As is to be expected, direct detection also becomes important for smaller $\mu$, with the latest bound from Xenon-1T~\cite{1705.06655} requiring $\mu\gtrsim800$\,GeV. 

Next, let us comment on the possibility of very large $\mu$. 
Recall that the $\tilde{B}$-$\tilde{\mu}_L$-$\tilde{\mu}_R$ contribution to the $g-2$ is proportional to the left-right mixing and is enhanced for large $\mu\tan\beta$. 
In this limit it is therefore, in principle, possible to explain the muon $g-2$ with very large slepton masses\footnote{Note that eventually $\Delta_\mu$ in Eq.~\eqref{eq:Delta} becomes large, and $m^2_{LR}$ approaches a maximal value.}. 
However, large $\mu\tan\beta$ will also eventually lead to charge-breaking minima in the the scalar potential~\cite{hep-ph/9507294, hep-ph/9612464, 1011.0260}; vacuum stability then leads to an upper limit on the smuon mass that can explain the $g-2$. 
This was explored in detail in Ref.~\cite{1309.3065}, where they obtained the bound $m_{\tilde{\mu}_1}\lesssim1.4\,(1.9)\,$TeV in order to account for the $g-2$ within $1\sigma\,(2\sigma)$.
This assumes that the stau is decoupled; however, note that non-universal slepton masses can also lead to potentially dangerous lepton flavour violation, and possibly CP violation~\cite{1309.3065}. 
On the other hand, many SUSY breaking scenarios assume universal slepton masses (at least at high scale). 
The vacuum stability bound is then significantly stronger due to the larger tau Yukawa, leading to $m_{\tilde{\mu}_1}\lesssim460\,$GeV~$(2\sigma)$~\cite{1309.3065}.
This already provides a very stringent constraint on this scenario in the case of universal slepton masses.

Turning now to the right panel of Fig.~\ref{fig:B-W}, we see that the situation is rather different for smaller values of $\tan\beta$. 
Since the SUSY contributions to the $g-2$ are proportional to $\tan\beta$, significantly lighter sleptons are required in order to explain the observed value. 
This means that LHC slepton searches~\cite{1803.02762, CMS-PAS-SUS-17-009} can already probe some of the best-fit region for the $g-2$; however, for small slepton masses these searches lose sensitivity due to the reduced mass difference between the sleptons and the LSP. 

A similar situation arises when one considers the effect of increasing the LSP mass. 
The regions consistent with direct detection and the $g-2$ remain qualitatively similar to those in Fig.~\ref{fig:B-W} (while always requiring $m_{\tilde{l}}>m_{\chi_1^0}$). 
On the other hand, the LHC slepton searches begin to lose their effectiveness as the spectrum becomes compressed. 
These fully compressed regions of parameter space are difficult to probe at the LHC, but could be tested by a future lepton collider.

In Fig.~\ref{fig:B-W} we also show the projected future reach of direct detection experiments and the (HL)-LHC. 
The vertical dashed line shows the reach of the LZ experiment~\cite{1802.06039}, which will be able to probe values of $\mu$ well beyond 1\,TeV. 
The horizontal dashed lines show the projected reach of the ATLAS slepton search with 300 or 3000\,$\text{fb}^{-1}$. 
This projection is obtained via a naive rescaling of the current expected limit by $\sqrt{\mathcal{L}}$.
We also assume that the limit on the cross-section remains constant when extrapolating to higher slepton masses. 
This is likely to result in a conservative estimate of the future reach, since with increased luminosity it will be possible to include additional dedicated signal regions for heavier sleptons. 
Overall, we find that the combination of LZ and the HL-LHC should be able to probe most of the region consistent with the $g-2$. 
Furthermore, the bounds on the wino-like NLSP from compressed searches can also be expected to improve in the future~\cite{1409.4533, 1804.05238}. 
A recent analysis suggests that soft lepton searches may eventually be able to exclude wino masses below 310\,(430)\,GeV with $300\,(3000)\,\text{fb}^{-1}$~\cite{1804.05238}. 
A complementary approach based on indirect, one-loop effects in di-lepton production at the HL-LHC~\cite{1711.05449}, or a $\sqrt{s}=250$\,GeV ILC~\cite{1504.03402}, also has a potential reach of around 300\,GeV.

One exception to the above conclusion is the very large $\mu\tan\beta$ region, where in principle the sleptons could be as heavy as 1.9\,TeV. 
To completely explore this region would require a future high-energy collider. 
Finally, one should also take into account that the uncertainty on the $g-2$ measurement will decrease significantly in the relatively near future. 
Assuming that the central value remains unchanged, much of the best-fit region may be probed at the LHC with a significantly lower integrated luminosity.

\subsection{$\tilde{B}-\tilde{\ell}$ co-annihilation}

\begin{figure*}[t]
  \includegraphics[width=0.45\textwidth]{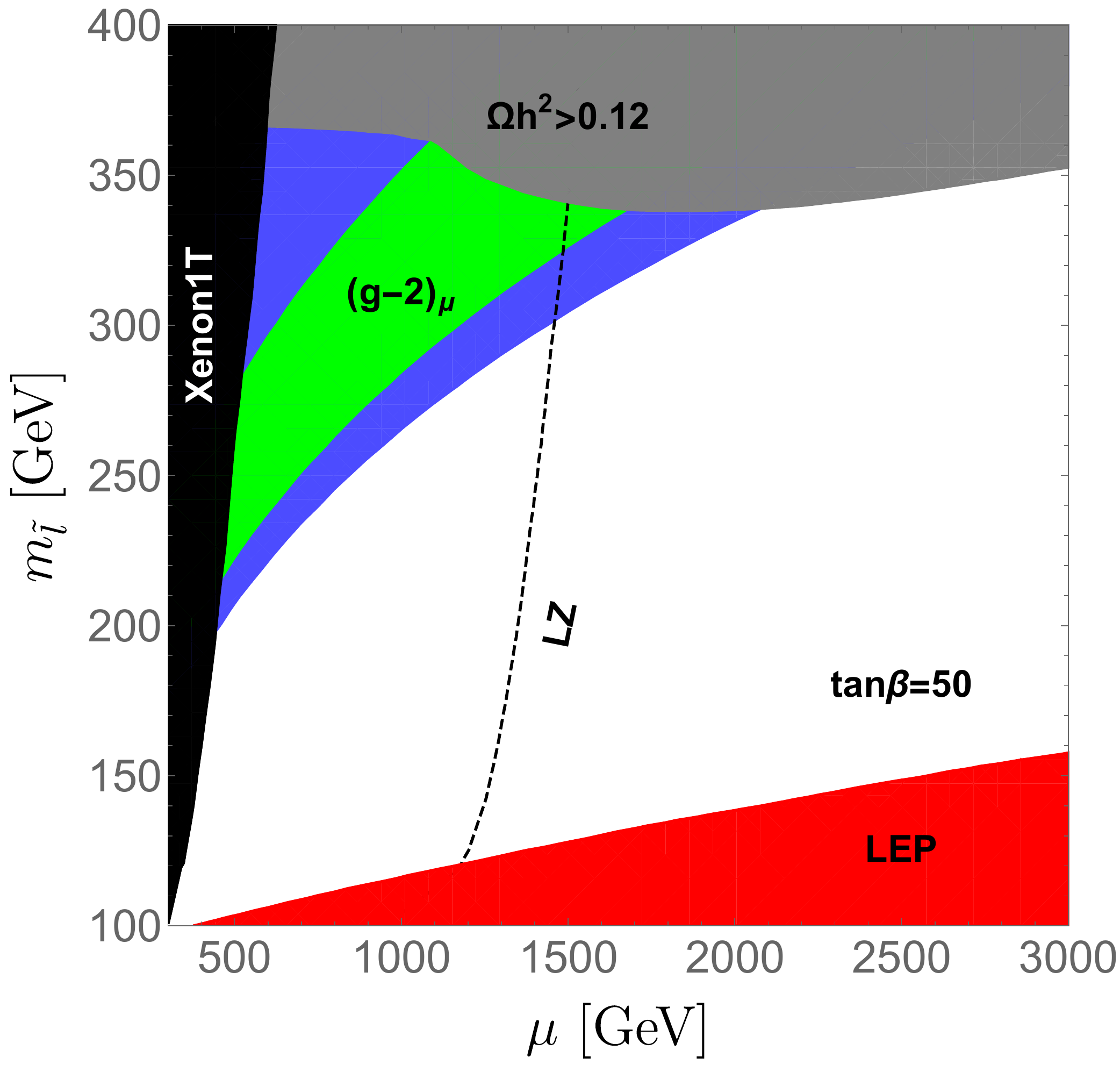}
  \includegraphics[width=0.45\textwidth]{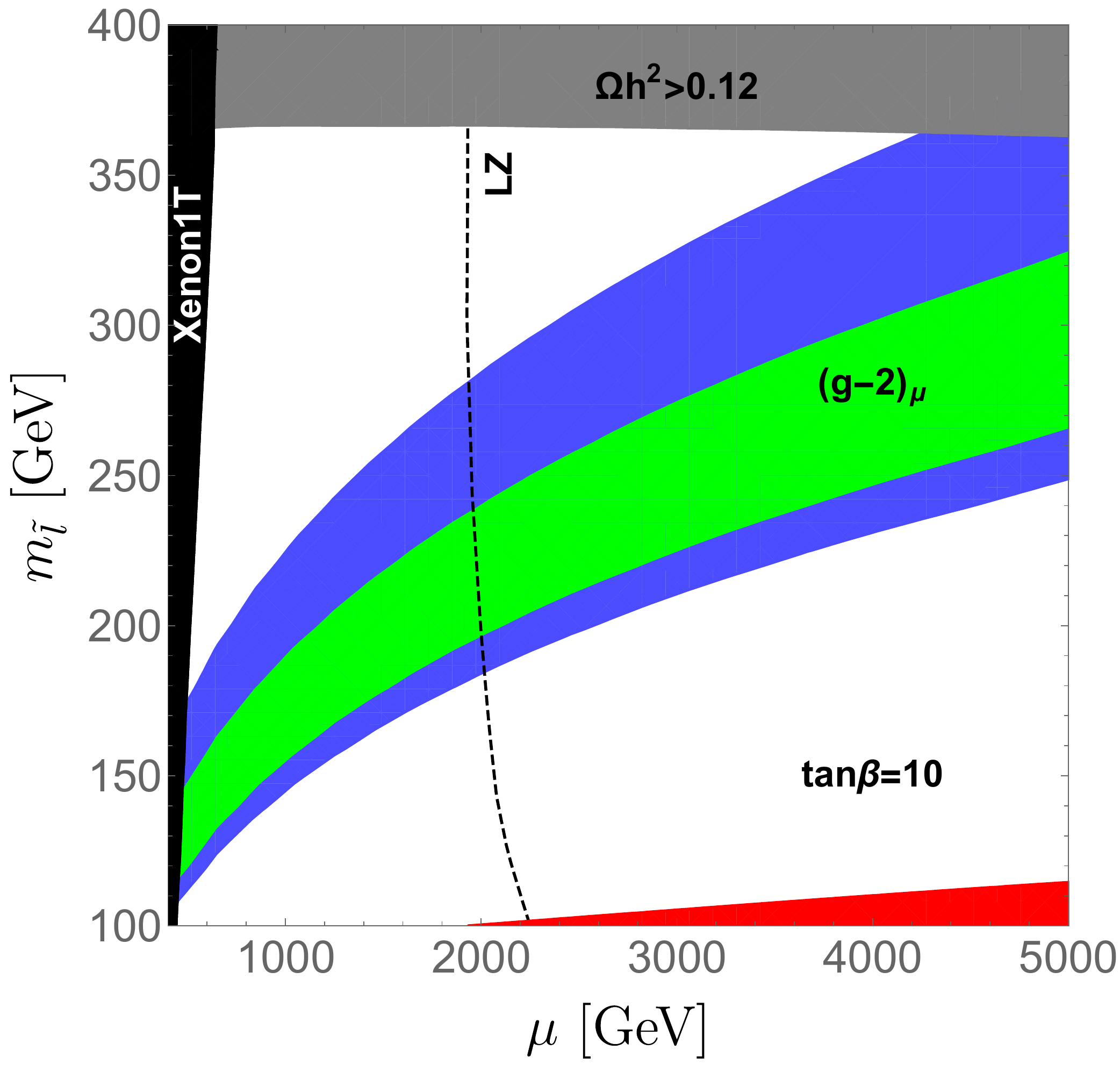}
  \caption{Current constraints and future projections for the case of $\tilde{B}-\tilde{l}$ co-annihilation. The green\,(blue) regions can explain the observed $(g-2)_\mu$ at $1\sigma\,(2\sigma)$. The black region is excluded by Xenon-1T~\cite{1705.06655}, while the dashed line denotes projected future reach. In the grey region it is not possible to obtain the correct DM density. The red region is excluded by LEP~\cite{LEP-slepton}. The left\,(right) panels are for $\tan\beta=50\,(10)$.} 
  \label{fig:B-l}
\end{figure*}

For the case of $\tilde{B}-\tilde{l}$ co-annihilation, a small mass splitting of $\lesssim10\,$GeV is required in order to achieve the correct DM relic density~\cite{hep-ph/0206266}.
However, the required mass-splitting falls quickly as the LSP mass increases, and the maximum DM mass that can be achieved in this scenario is generally quite low. 
This is in contrast to the case of $\tilde{B}-\tilde{W}$ co-annihilation where the correct abundance can be achieved for masses all the way up to $\sim3\,$TeV. 
The detailed phenomenology also depends on the identity of the lightest slepton. 
If one assumes universal soft masses for the sleptons then this is likely to be the lightest stau; however, this particular case is already highly constrained, as we shall discuss later. 
Therefore, we will predominantly focus on the minimal scenario where only the first and second generation sleptons are light and involved in the co-annihilation\footnote{A UV model which realises this spectrum will appear in a forthcoming paper~\cite{newmodel}.}.
In this case we find that the DM and slepton masses should be below around 350\,GeV in order to obtain the correct relic abundance\footnote{It is possible to increase the LSP mass by considering very large $\mu\tan\beta$, where the left-right mixing becomes important.}, which will provide an important constraint on the viable parameter space. 

In Fig.~\ref{fig:B-l} we show the best-fit region to explain the muon $g-2$, along with the various experimental constraints. 
We have assumed that the wino is decoupled ($M_2=3$\,TeV), while the bino mass has been adjusted to obtain the correct DM abundance through $\tilde{B}-\tilde{l}$ co-annihilation. 
Firstly, notice the strong upper bound on $m_{\tilde{l}}$, above which it is no longer possible to obtain the correct relic density. 
It is worth highlighting that the NLSP, and hence dominant co-annihilation partner, varies across the parameter space: at small $\mu\tan\beta$ it is the sneutrinos, and then eventually becomes the lightest smuon. 
This transition is visible around $\mu=1\,$TeV in the left panel of Fig.~\ref{fig:B-l}. 

The compressed spectrum makes this scenario very challenging to probe at the LHC. 
Although there are dedicated slepton searches for compressed spectra~\cite{1712.08119}, these currently only have sensitivity to slepton masses up to 190 GeV for $\Delta m=5\,$GeV, with the reach decreasing rapidly for smaller or larger mass-splittings. 
Furthermore, this assumes all of the first and second generation sleptons are degenerate and have a small mass splitting to the LSP. 
Across much of the parameter space this is not the case, and the compressed slepton search is then only sensitive to the lightest charged slepton, significantly reducing the production cross-section. 
Consequently, this search does not provide any constraint on the region that can explain the muon $g-2$. 
It is also likely that this will remain the case even at the HL-LHC; however a detailed analysis is beyond the scope of this paper. 

Direct detection therefore provides the best prospects for testing this scenario in the near future.  
The LZ experiment will be able to probe a significant fraction of the viable parameter space, especially for large $\tan\beta$. 
Nevertheless, there remain significant regions of parameter space that cannot be tested by direct detection experiments. 
On a more positive note, obtaining the correct DM abundance via co-annihilation requires light sleptons; these could be discovered or excluded by a future 1\,TeV lepton collider.  

Finally, let us briefly comment on some extensions to this minimal scenario in which other particles may also be relatively light. 
If the wino mass is not too large, it can provide additional contributions to the muon $g-2$.  
However, these contributions only dominate in the small $\mu$ region which is anyway constrained by direct detection; the situation is then similar to that shown in Fig.~\ref{fig:B-l}. 
Depending on its mass, such a wino may also be visible at the LHC~\cite{1801.03957, 1803.02762}.
Another well-motivated possibility is that of universal slepton masses.
In this case the lightest stau will generally be the NLSP and play the dominant role in the co-annihilation. 
This leads to a significantly weaker upper bound on $m_{\tilde{l}}$ from the relic density.   
However, one must instead take into account the vacuum stability bound discussed previously, which significantly restricts the parameter space~\cite{1309.3065}. 
Lastly, throughout our analysis we have assumed $m_{\tilde{l}_L}=m_{\tilde{l}_R}$, however it is possible that only the left- or right-handed sleptons may be light. 
In this case the DM abundance can still be obtained via $\tilde{B}-\tilde{l}$ co-annihilation, but light higgsinos are required in order to explain the muon $g-2$.
This scenario was explored in detail in Ref.~\cite{1704.05287}, where they found that it can be covered by next-generation direct detection experiments and the HL-LHC. 

\section{Conclusion}
A simultaneous explanation of both the observed DM abundance and the muon $g-2$ remains a viable and interesting possibility within the MSSM. 
We have argued that in light of the stringent bounds from direct detection, the most promising scenario involves bino-like DM produced via either bino-wino or bino-slepton co-annihilation. 
In the former case, much of the interesting parameter space can be tested in the future with next-generation direct detection experiments and the HL-LHC. 
However, in the most extreme case of very large $\mu$, the sleptons could be as heavy as 1.9\,TeV; a new high-energy collider is then needed to cover the full parameter space. 
The bino-slepton co-annihilation case is extremely difficult to test at the LHC, but can be partially probed by future direct detection experiments. 
Furthermore, obtaining the correct relic abundance requires light sleptons below around 350\,GeV, within reach of a future 1\,TeV lepton collider. 

\begin{acknowledgments}
This work is supported by Grants-in-Aid for Scientific Research from the Ministry of Education, Culture, Sports, Science, and Technology (MEXT), Japan, No. 26104001 (T.T.Y.), No. 26104009 (T.T.Y.), No. 16H02176 (T.T.Y.) and No. 17H02878 (T.T.Y.), and by the World Premier International Research Center Initiative (WPI), MEXT, Japan (P.C., C.H. and T.T.Y.). 
\end{acknowledgments}

\bibliographystyle{apsrev4-1-mod.bst}
\bibliography{gm2DM}

\end{document}